\documentclass[]{emulateapj}

\usepackage{graphicx,times}
\slugcomment{Submitted 2014 February 28; 2014 Accepted April 19}
\shorttitle{The effect of the turbulent kinetic flux}
\shortauthors{Q. S. Zhang}

\begin{document}

\title{The solar abundance problem: the effect of the turbulent kinetic flux on the solar envelope model}
\author{Q. S. Zhang \altaffilmark{1,2}}
\email{zqs@ynao.ac.cn(QSZ)}

\altaffiltext{1}{Yunnan Observatories, Chinese Academy of Sciences, P.O. Box 110, Kunming 650011, China.}
\altaffiltext{2}{Key Laboratory for the Structure and Evolution of Celestial Objects, Chinese Academy of Sciences, Kunming, 650011, China.}

\begin{abstract}

Recent 3D-simulations have shown that the turbulent kinetic flux (TKF) is significant. We discuss the effects of TKF on the size of convection zone and find that the TKF may help to solve the solar abundance problem.
The solar abundance problem is that, with new abundances, the solar convection zone depth, sound speed in the radiative interior, the helium abundance and density in the convective envelope are not in agreement with helioseismic inversions.
We have done Monte Carlo simulations on solar convective envelope models with different profile of TKF to test the effects. The solar abundance problem is revealed in the standard solar convective envelope model with AGSS09 composition, which shows significant differences ($\rm{\sim 10 \%}$) on density from the helioseicmic inversions, but the differences in the model with old composition GN93 is small ($\rm{\sim 0.5 \%}$). In the testing models with different imposed TKF, it is found that the density profile is sensitive to the value of TKF at the base of convective envelope and insensitive to the structure of TKF in the convection zone. Required value of turbulent kinetic luminosity at the base is about $\rm{-13\%\sim-19\%L_{\odot}}$. Comparing with the 3D-simulations, this value is plausible. This study is for the solar convective envelope only. The evolutionary solar models with TKF are required for investigating its effects on the solar interior structure below the convection zone and the whole solar abundance problem, but the profile of TKF in the overshoot region is needed.

\end{abstract}


\section{Introduction}
\label{sec1}

Recent photospheric analysis (e.g., \citet{AGS05,AGSS09,caf11}) have indicated that the solar photospheric metallicity is significantly lower than
the old values \citep{GN93,GS98}. This arises the solar abundance problem that standard solar models with the revised composition show serious deviations from the helioseismic inversions, i.e., the depth of the convection zone (CZ) $\rm{r_{bc}}$, the surface helium abundance $\rm{Y}$ and
the sound speed and density in the solar interior \citep{ba04,bah05,bah06,yb07}. Many models are proposed to modify the solar model, including enhanced diffusion \citep{guz05}, the accretion model \citep{guz05,guz10,ser11}, particles with an axion-like interaction \citep{Vincent13}, etc.. However, none has succeeded in solving the problem. The accretion model have shown improvements on the solar model, but the inconsistence remains since $\rm{r_{bc}}$ and $\rm{Y}$ can not fit the helioseismic restrictions simultaneously \citep{guz05,guz10,ser11}. It is expected that the opacity at the base of the convection zone (BCZ) could be adjusted upward, since the tests of enlarged opacity reduce the discrepancies between model and helioseismic inversions \citep{ba04,bah05,chr09}.

The turbulent kinetic flux (TKF) is usually ignored in modeling star because it is thought to be small. Another reason is that there is no widely accepted nonlocal convection theory to describe the TKF. However, 3D-simulations on stellar convection envelopes have shown that TKF is not ignorable in some cases \citep{tian09,hotta14}. It should be noticed that the TKF at BCZ is negative because the turbulent convection transport kinetic energy from the CZ to the radiative region. A negative TKF requires radiative flux plus convective flux being larger, thus leads to a deeper convective boundary. This could help to improve the solar model. Based on this ideal, we test the TKF in solar convective envelope (CE) models.

\section{Effects of turbulent kinetic flux (TKF) on convective boundary}
\label{sec2}

The TKF is the transport of turbulent kinetic, which is a non-local effect of turbulent convection. When TKF is taken into account, the stellar energy equation is as follows:
\begin{eqnarray} \label{energyeq}
\frac{{\partial [4\pi {r^2}({F_R} + {F_C} + {F_K})]}}{{\partial {m_r}}} = {\varepsilon _N} - {\varepsilon _{neu}} + {\varepsilon _g},
\end{eqnarray}%
where $\rm{F_R}$, $\rm{F_C}$ and $\rm{F_K}$ are the radiative flux, the convective flux and the TKF, respectively, and $\rm{\varepsilon _N}$, $\rm{\varepsilon _{neu}}$ and $\rm{\varepsilon _g}$ are the energy contributed by the nuclear burning, the neutrino loss and the entropy variation, respectively. Their sum is the total flux $\rm{F}$. According to the turbulent theories (e.g., \citet{xiong81,xiong85,li07,ma10}), TKF satisfies the following equation:
\begin{eqnarray} \label{turbkeq}
\frac{{\partial (4\pi {r^2}{F_K})}}{{\partial {m_r}}} = \frac{{\delta g{F_C}}}{{\rho {c_P}T}} - {\varepsilon _{turb}},
\end{eqnarray}%
where $\rm{\varepsilon _{turb}}$ is the turbulent dissipation, the first term in the r.h.s. is the turbulent production caused by buoyancy working.
Accordingly, the stellar energy equation can be rewritten as:
\begin{eqnarray} \label{energyeq2}
\frac{{\partial [4\pi {r^2}({F_R} + {F_C})]}}{{\partial {m_r}}} = {\varepsilon _N} - {\varepsilon _{neu}} + {\varepsilon _g} + {\varepsilon _{turb}} - \frac{{\delta g{F_C}}}{{\rho {c_P}T}},
\end{eqnarray}%
which shows the processes of the turbulent dissipation transfers the kinetic energy to the thermal energy and buoyancy transfers the thermal potential energy to the kinetic energy. Equations (\ref{energyeq}), (\ref{turbkeq}) and (\ref{energyeq2}) are the equations of total energy, the kinetic energy and the thermal energy, respectively.

The TKF $\rm{F_K}$ should be determined by non-local convection theory. In local convection theory, e.g., the widely used mixing lenth theory (MLT), the turbulence is assumed to be in local equilibrium thus there is always $\rm{F_K=0}$. However, the non-local convection theories are difficult so that there are few information known at present. On the other hand, recent studies have shown that TKF seems to be far away from ignorable because it is comparable with the total flux in some cases.
\citet{tian09} have shown that TKF is as large as $\rm{\sim10\%}$ or more of the negative total flux in their simulations of downward overshoot in RGB stars (see Fig.2 in their paper). \citet{hotta14} have shown that, in the most part of the solar CZ, TKF is comparable with the total flux, i.e., $\rm{F_{k}\sim-F_{tot}}$. However, in the simulations by \citet{hotta14}, the real value of $\rm{F_K}$ at the BCZ can not be revealed since the bottom boundary was set at the BCZ.
Here, we simply discuss the effects of TKF on the size of CZ.

The total flux comprises three parts, thus,
\begin{eqnarray} \label{flux}
\frac{{\lambda T}}{{{H_P}}}{\nabla _R} = F = {F_R} + {F_C} + {F_K} = \frac{{\lambda T}}{{{H_P}}}\nabla  + {F_C} + {F_K},
\end{eqnarray}%
where $\rm{\nabla _R}$ is the radiative temperature gradient, $\rm{\nabla}$ is the temperature gradient in stellar interior.
In the CZ in stellar interior (e.g., $\rm{lgT>6}$), $\rm{\nabla\approx \nabla_{ad}}$ can be ensured because of high P\'{e}clet number. Therefore the convective flux satisfies:
\begin{eqnarray} \label{convflux}
{F_C} \approx \frac{{\lambda T}}{{{H_P}}}({\nabla _R} - {\nabla _{ad}} - \frac{{{H_P}{F_K}}}{{\lambda T}}).
\end{eqnarray}%
The CZ is defined by $\rm{F_C>0}$, which means the buoyancy works on fluid elements also the convective instability. Accordingly, the convective criterion is:
\begin{eqnarray} \label{convcrit}
{\nabla _R} - {\nabla _{ad}} - \frac{{{H_P}{F_K}}}{{\lambda T}} > 0.
\end{eqnarray}%
At the boundaries of CZ, turbulent flows transport turbulent kinetic energy from the CZ to the overshoot region. Therefore $\rm{F_K>0}$ in the top boundary of CZ (e.g., the convective core boundary) and $\rm{F_K<0}$ in the BCZ. In the case of convective core, $\rm{F_K>0}$ leads to a deeper convective boundary and a small convective core comparing with the Schwarzschild criterion. In the case of stellar CE, $\rm{F_K<0}$ leads to a deeper convective boundary and a larger CE comparing with the Schwarzschild criterion.

This effect may helps to solve the solar abundance problem. In standard solar model with revised composition \citep{AGSS09}, the BCZ is too shallow to fit the helioseismic value, since the metallicity is low. In order to enlarge the CE, the opacity at the BCZ is expected to be larger. It has been found that a required upward adjustment on the opacity is about $\rm{10\% - 30\%}$ \citep{ba04,bah05,chr09}. It should be noticed that taking into account TKF leads to similar effect as increasing the opacity. It is convenient to define another radiative temperature gradient for thermal flux:
\begin{eqnarray} \label{convcrit}
{\nabla _{R,Therm}} = \frac{{{H_P}}(F_R+F_C)}{{\lambda T}} = {\nabla _{R}} - \frac{{{H_P}{F_K}}}{{\lambda T}},
\end{eqnarray}%
which describes the required temperature gradient if the thermal energy is transported by only radiation. The convective criterion is $\rm{{\nabla _{R,Therm}}>{\nabla _{ad}}}$ and it is not difficult to find:
\begin{eqnarray} \label{convcrit}
{\nabla _{R,Therm}} \propto \kappa ({F_C} + {F_R}) \propto \kappa (1 - \frac{{{F_K}}}{F}).
\end{eqnarray}%
Therefor a negative TKF results in a similar effect as increasing the opacity.

\citet{arnett10} have stated that some discrepancies between the standard solar model and helioseismic inversions may be caused by ignoring some significant aspects of convection which includes the TKF. Arnett et al. have proposed that the internal gravity wave excited at the BCZ transports energy inward and changes the solar structure as the similar way as increasing the opacity \citep{guz06,guz10}. This effects is similar to the TKF because the gravity wave below the BCZ also shows a negative energy flux.

\section{The model of the present Solar convective envelope (CE)}
\label{sec3}

To study the solar CE is easier than to study the solar evolutionary models.
The solar CE is homogeneous and the composition has been determined via observations and helioseismic technic. The luminosity in the solar CE can be regarded as a constant, since there is no nuclear burning significantly releases energy and the gravity energy is ignorable for the main-sequence sun. Additionally, the solar radius and the solar luminosity are also determined. As a consequence, the structure of the present solar CE satisfies the differential equations with initial conditions as follows:
\begin{eqnarray} \label{dynamic}
\frac{{d(P + {P_{turb}})}}{{dr}} =  - \rho g,
\end{eqnarray}%
\begin{eqnarray} \label{densitydef}
\frac{{d{m_r}}}{{dr}} = 4\pi {r^2}\rho ,
\end{eqnarray}%
\begin{eqnarray} \label{grt}
\frac{{d\ln T}}{{d\ln P}} = \nabla ,
\end{eqnarray}%
\begin{eqnarray} \label{eos}
P = P(\rho ,T,Y,Z),
\end{eqnarray}%
\begin{eqnarray} \label{constantL}
4\pi {R_{\odot}^2}\sigma {T_S^4} = L_{\odot}= L_r= 4\pi {r^2}({F_C} + {F_R} + {F_K}),
\end{eqnarray}%
\begin{eqnarray} \label{solarmass}
{m_s} = M_{\odot},
\end{eqnarray}%
and
\begin{eqnarray} \label{atm}
\frac{{d\ln \rho }}{{d\tau }} = \delta [\frac{g}{{P\kappa }}{(\frac{{\partial \ln T}}{{\partial \ln P}})_\rho } - \frac{{d\ln T}}{{d\tau }}].
\end{eqnarray}%
Equation (\ref{eos}) is the equation of state and equation (\ref{atm}) is the integral of the atmosphere determining the density at the surface $\rm{\rho_S}$. The K-S $\rm{T-\tau}$ relation \citep{ks66} is adopted in equation (\ref{atm}). Equations (\ref{constantL}), (\ref{solarmass}) and the density at the surface determined by equation (\ref{atm}) are three initial conditions. $\rm{P_{turb}=\rho \overline{u_r^2}}$ is the turbulent pressure where $\rm{\overline{u_r^2}}$ represents the radial kinetic energy estimated by the adopted convection theory. $\rm{\nabla}$ is the temperature gradient determined by adopted convection theory.
An advantage of study the solar CE is that we can exclude the uncertainties in modeling the radiative core and the chemical evolutionary history.

Calculating the properties of the nonlocal convection via nonlocal convection theories is difficult, and there is even no widely accepted nonlocal convection theory. For this sake, the TKF $\rm{F_K}$ is imposed in external in our calculations to test the effects. And the temperature gradient is calculated via the local convection theory MLT. Both local and nonlocal convection theories show similar result that the solar CE is in adiabatical stratification in most region.

The imposed TKF $\rm{F_K}(=L_K/(4\pi r^2))$ is as follows:
\begin{eqnarray} \label{impFk}
L_K = \left\{ {\begin{array}{*{20}{c}}
   {{L_{K,bc}} + ({L_{K,cz}} - {L_{K,bc}})\frac{{r - {r_{bc}}}}{{{r_0}}},({r_{bc}} \le r \le {r_{bc}} + {r_0})}  \\
   {{L_{K,cz}},(r \ge {r_{bc}} + {r_0},\lg T \ge b)}  \\
   {{L_{K,S}} + ({L_{K,cz}} - {L_{K,S}})\frac{{\lg T - a}}{{b - a}},(a \le \lg T \le b)}  \\
   {{L_{K,S}},(\lg T \le a)}  \\
\end{array}} \right.
\end{eqnarray}%
where $\rm{{L_{K,bc}}}$, $\rm{{L_{K,cz}}}$, $\rm{{L_{K,S}}}$, $\rm{r_0}$, $\rm{a}$ and $\rm{b}$ are parameters. The motivations of those parameters are as follows. $\rm{{L_{K,bc}}}$ and $\rm{{L_{K,S}}}$ provide energy for the downward and upward overshoot. $\rm{{L_{K,cz}}}$ represents the average kinetic luminosity in the most range of the CZ, motivated by the simulation \citep{hotta14} which shows the kinetic luminosity being approximate a constant ($\rm{L_{K,cz} \sim - L_{sun}}$) in the most part of the CZ. The swap region between $\rm{{L_{K,bc}}}$ and $\rm{{L_{K,cz}}}$ with the length $\rm{r_0}$ near the BCZ is set because the turbulent diffusion and the dissipation dominates at here. The swap region between $\rm{{L_{K,S}}}$ and $\rm{{L_{K,cz}}}$ between $\rm{a<lgT<b}$ (typically we adopt $\rm{a \leq 3.9 \leq b}$) is set because the most violent turbulence is near the temperature about $\rm{lgT=3.9}$ and the turbulent diffusion varieties fast at here.

We integrate equations (\ref{dynamic}), (\ref{densitydef}) and (\ref{grt}) from the surface downward to the BCZ (i.e., where $\rm{{\nabla _{R,Therm}}={\nabla _{ad}}}$) with corresponding chemical composition. We do not integrate equations into the overshoot region because the chemical abundance is unknown. The overshoot mixing is of low efficiency \citep{zha13} thus the chemical abundance gradient caused by the settling may not be removed. The MLT parameter $\rm{\alpha}$ is iteratively adjusted to ensure $\rm{r_{bc}}$ being consistent with the helioseimic inversion. The equation of state is interpolated from OPAL-EOS tables \citep{EOS2005}, the opacity is interpolated from OPAL tables \citep{opacity} with corresponding metal compositions and the low temperature opacity table by \citet{F05}. The solar CE models calculated based on the above scheme are of correct helium abundance and $\rm{r_{bc}}$. We need to compare the sound speed and the density with the helioseismic inversions to check whether the envelope models are consistence with the helioseismic restrictions.

\section{Numerical results}
\label{sec4}

\begin{figure*}
\centering
\includegraphics[scale=1.2]{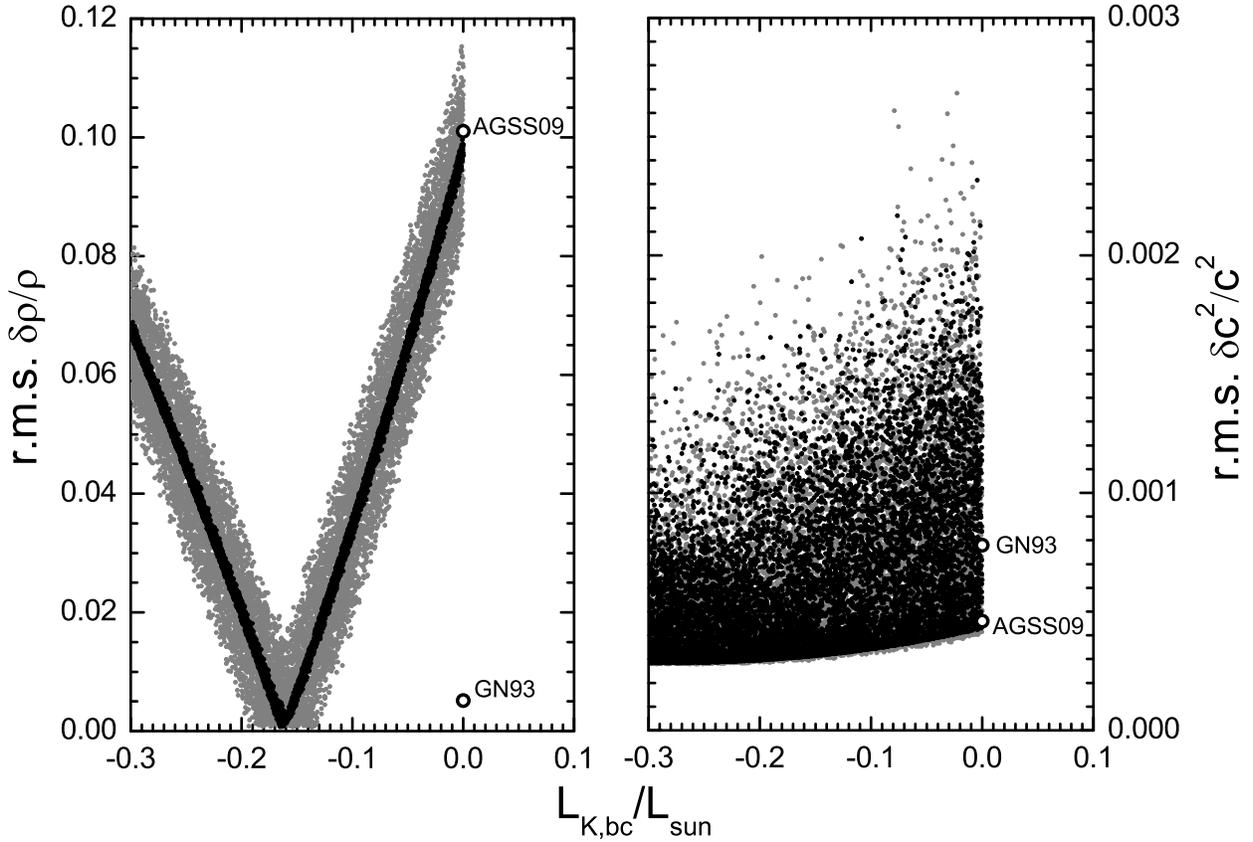}
\caption{Monte Carlo simulation of solar CE models with revised composition \citep{AGSS09}. Black points ($\rm{10000}$ models) are in uniform distribution for $\rm{-0.3 \leq L_{K,bc}/L_{\odot} \leq 0 }$, $\rm{-1.5 \leq L_{K,cz}/L_{\odot} \leq 0 }$, $\rm{0 \leq L_{K,s}/L_{\odot} \leq 0.1 }$, $\rm{0.05 \leq r_0/R_{\odot} \leq 0.1 }$, $\rm{3.76 \leq a \leq 3.9 }$, $\rm{3.9 \leq b \leq 4.1 }$ and the factor of $\rm{P_{turb}}$ between 0.5 and 2.0, with $\rm{Y=0.2485}$ and $\rm{r_{bc}/R_{\odot}=0.7135}$. Gray points (about $\rm{11000}$ models) are also in uniform distribution with the parameter spaces are same as black points, but with uncertainties of $\rm{0.2455 \leq Y \leq 0.2515 }$ \citep{ba04} and $\rm{0.713 \leq r_{bc}/R_{\odot} \leq 0.714 }$ \citep{ba98}. Standard solar CE models GN93 (with old composition \citep{GN93}) and AGSS09 are also shown by the empty circles. }\label{rms}
\end{figure*}

We have done Monte Carlo simulations with about 21000 solar CE models. The results are shown in Fig.\ref{rms}. The helioseismic inversions density and sound speed are referenced from \citet{ba09}. Two kinds of models are calculated: the black points with fixed $\rm{r_{bc}}$ and $\rm{Y}$ and in uniform distribution of parameters of the TKF and the factor of $\rm{P_{turb}}$, and the gray points additionally taking into account the uncertainties of $\rm{r_{bc}}$ and $\rm{Y}$. The parameter spaces are described in the table comments. Two standard CE models with out TKF and turbulent pressure, GN93 and AGSS09, are also shown for comparison. It is found that, with the helioseismic $\rm{r_{bc}}$ and $\rm{Y}$, the standard CE model AGSS09 with revised composition \citep{AGSS09} show significant inconsistence (about $\rm{10\%}$) on the density. The GN93 CE model with old composition \citep{GN93} seems much better than the AGSS09 model, since the r.m.s. difference of density is about $\rm{0.5\%}$. The inconsistence of AGSS09 CE model indicates that, when the standard stellar structure equations and standard input physics (e.g., equation of state, opacity and etc.) are used in the solar CE, it is impossible to obtain a solar model with AGSS09 composition fitting all helioseismic restrictions whatever modifications are adopted in the solar radiative core or in the chemical evolution equations. This may be the reason of why $\rm{r_{bc}}$ and $\rm{Y_S}$ can not fit the helioseismic restrictions simultaneously in the accretion model \citep{guz05,ser11}.

\begin{table}
\centering
\caption{ Correlation coefficients between r.m.s. errors of sound speed / density and parameters of the solar CE models. }\label{cor}
\begin{tabular}{lcccc}
\hline\hline\noalign{\smallskip}
r.m.s. of  & $\rm{\delta c^2 / c^2}$ & $\rm{\delta \rho / \rho}$ & $\rm{(\delta \rho / \rho)_A}$ & $\rm{(\delta \rho / \rho)_B}$ \\
\hline\noalign{\smallskip}
$\rm{{L_{K,bc}}}$    &  $\rm{0.3636}$   & $\rm{0.3632}$    & $\rm{0.9645}$    & $\rm{-0.9392}$  \\
$\rm{{L_{K,cz}}}$    & $\rm{-0.3765}$   & $\rm{-0.01403}$  & $\rm{-0.003777}$ & $\rm{-0.04726}$ \\
$\rm{{L_{K,S}}}$     & $\rm{-0.05255}$  & $\rm{0.004576}$  & $\rm{0.01702}$   & $\rm{-0.01722}$ \\
$\rm{r_0}$           & $\rm{0.001676}$  & $\rm{0.002756}$  & $\rm{0.02304}$   & $\rm{-0.02821}$ \\
$\rm{a}$             & $\rm{-0.1559}$   & $\rm{0.01137}$   & $\rm{0.0006198}$ & $\rm{0.02404}$  \\
$\rm{b}$             & $\rm{-0.1369}$   & $\rm{0.009944}$  & $\rm{0.009527}$  & $\rm{0.01193}$  \\
$\rm{P_{turb}}$      & $\rm{0.5704 }$   & $\rm{-0.001292}$ & $\rm{-0.01805}$  & $\rm{0.02913}$  \\
$\rm{Y}$             & $\rm{0.1031 }$   & $\rm{0.02819}$   & $\rm{0.2162}$    & $\rm{-0.3145}$  \\
$\rm{r_{bc}}$        & $\rm{-0.04167}$  & $\rm{-0.02319}$  & $\rm{-0.1158}$   & $\rm{0.1406}$   \\
\hline\hline
\end{tabular}
\tablecomments{ The correlation coefficient between row name and column name are shown in corresponding row and column. The samples are the gray points in Fig.\ref{rms}. The column $\rm{(\delta \rho / \rho)_A}$ is for $\rm{{L_{K,bc}>-0.164L_{sun}}}$ (about $\rm{6000}$ models), and the column $\rm{(\delta \rho / \rho)_B}$ is for $\rm{{L_{K,bc}<-0.164L_{sun}}}$ (about $\rm{5000}$ models). This split is for the reason that, basically, the density in the CE models is higher than the helioseismic inversions for $\rm{{L_{K,bc}>-0.164L_{sun}}}$ and lower for $\rm{{L_{K,bc}<-0.164L_{sun}}}$. The r.m.s. errors are absolute value thus its tendency changes near $\rm{{L_{K,bc}=-0.164L_{sun}}}$. In this case, doing statistics respectively is more reasonable. }
\end{table}

The correlation coefficients for r.m.s. differences of density and sound speed are listed in Table \ref{cor}. It is shown that some parameters can affect the sound speed. However, the resulting sound speeds are always consistent with the helioseismic inversions ($\rm{c^2}$ in the accuracy of $\rm{0.2\%}$) as shown in Fig.\ref{rms}, and the sound speed differences are also affected by the $\rm{T-\tau}$ relation in the atmosphere and the interpolations of EOS. Therefore, we can't give meaningful 'best' value ranges of the parameters based on the sound speed differences.

\begin{figure}
\centering
\includegraphics[scale=0.6]{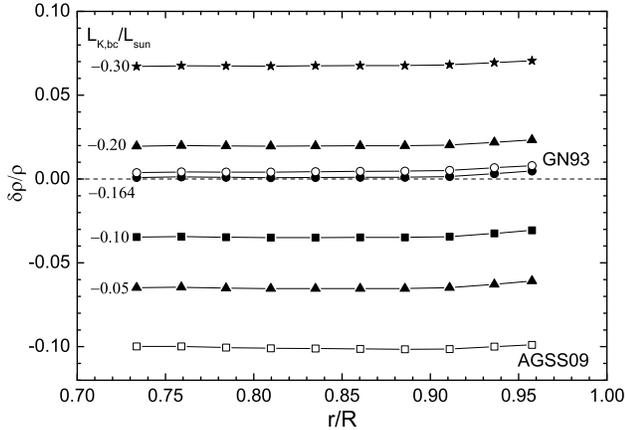}
\caption{Differences of density between the solar CE models and the helioseismic inversions, where $\rm{\delta \rho / \rho=(\rho_{helio}-\rho_{model})/\rho_{helio}}$. The standard model GN93 and AGSS09 are shown in empty symbols. CE models with different $\rm{{L_{K,bc}}}$ are shown in full symbols, with other parameters: $\rm{{L_{K,cz}=-1.00L_{\odot}}}$, $\rm{{L_{K,S}=0.05L_{\odot}}}$, $\rm{r_0=0.05R}$, $\rm{a=3.8}$, $\rm{b=4.0}$ and the factor of $\rm{P_{turb}}$ being 1, with $\rm{Y=0.2485}$ and $\rm{r_{bc}/R_{\odot}=0.7135}$. }\label{rho}
\end{figure}

On the other hand, the correlation coefficients for the density show significant dependency on $\rm{L_{K,bc}}$, and are insensitive to other parameters of TKF. The affects of the value of $\rm{L_{K,bc}}$ on the density are shown in Fig.\ref{rho}. It is found that low $\rm{L_{K,bc}}$ results in low density in the solar CE. We can estimate $\rm{L_{K,bc}}$ from Fig.\ref{rms} and Fig.\ref{rho}: for $\rm{Y=0.2485}$ and $\rm{r_{bc}=0.7135 R_{\odot}}$, best value is about $\rm{L_{K,bc} =-0.164L_{\odot}}$, or, at least, $\rm{L_{K,bc}}$ should be in the range of $\rm{(-0.19L_{\odot}, -0.13L_{\odot})}$ taking into account the uncertainties of $\rm{Y}$ and $\rm{r_{bc}}$. However, the other parameters can't be estimated. It can be found by comparing the gray points with the black points in Fig.\ref{rms} that the uncertainties of $\rm{Y}$ and $\rm{r_{bc}}$ lead to significant dispersion. For fixed $\rm{L_{K,bc}}$, the black points show that the other parameters lead to a dispersion about $\rm{0.5\%}$ on r.m.s. error of density, and the gray points which takes into account the uncertainties of $\rm{Y}$ and $\rm{r_{bc}}$ show a dispersion about $\rm{3\%}$. This indicates that the uncertainties of $\rm{Y}$ and $\rm{r_{bc}}$ lead to a dispersion about $\rm{2.5\%}$, which is much larger than the dispersion caused by the parameters of TKF except $\rm{L_{K,bc}}$, thus the uncertainties of $\rm{Y}$ and $\rm{r_{bc}}$ prevent us to estimate the best values of other parameters. In a word, $\rm{{L_{K,bc}}}$ is the only sensitive parameter of TKF to the density of the CE. The reason of other parameters being insensitive may be that, the CE is in adiabatical stratification in most region thus the shape of $L_K$ and the turbulent pressure can hardly affect the structure in most region of the CE. The turbulent pressure is important only in a thin layer near the solar surface, i.e., $\rm{P_{turb}/P \approx 15\%}$ at $\rm{lgT \approx 3.9}$ but less than $\rm{0.1\%}$ in $\rm{r< 0.995R}$. There is no data in Basu et al.'s (2009) helioseismic inversions in $\rm{r> 0.995R}$ to detect the turbulent pressure.

\section{Conclusions and discussions}
\label{sec5}

In this letter, we discussed the effects of the turbulent kinetic flux (TKF) on the size of convection zone (CZ) and test the effects of TKF on the solar convective envelope (CE) models. The main conclusions are as follows:

(i) The presence of TKF modifies the convective criterion and makes convective boundaries shift downward, thus the convective core becomes smaller and the CE becomes larger.

(ii) The solar abundance problem is revealed in the solar CE models. The standard solar CE model with revised composition \citep{AGSS09} shows significant difference ($\rm{\sim 10\%}$) on density from the helioseismic inversions. This makes it be impossible to obtain a solar model with AGSS09 composition fitting all helioseismic restrictions if the standard stellar structure equations and standard input physics are used in the solar CE.

(iii) Taking into account TKF could improve the solar CE model. The density structure of the solar CE is sensitive to the value of the TKF at the BCZ and insensitive to its profile in the CZ. Required turbulent kinetic luminosity at the BCZ is $\rm{-13\% L_{\odot} < L_{K,bc} < -19\% L_{\odot} }$ taking into account the uncertainties of $\rm{Y}$ and $\rm{r_{bc}}$.

In this paper, we have done a limited test with the TKF on the solar abundance problem, i.e., testing required TKF to construct the convective envelope of the solar model (with AGSS09 composition) fitting all helioseismic restrictions. However, we can't show the effect of the TKF in complete solar evolutionary models. The TKF profile below the BCZ is required to do that. However, there is no simulation showing the profile. The standard model with AGSS09 abundance show lower helium abundance and shallower BCZ comparing with the helioseismic inversions. \citet{guz05} have shown that the sound speed differences below the BCZ can not be removed in some non-standard models even with the correct BCZ. We think that taking into account some important aspects of convection (the overshoot mixing and turbulent kinetic flux) could help to solve the solar abundance problem. The incomplete mixing caused by the convective overshoot \citep{zha13} partially compensates the settling thus enlarge the helium abundance. And the incomplete mixing below the BCZ is favored by the sound speed when the BCZ is in the correct location \citep{brun99,zha12,zha13}. A fault of the incomplete mixing is to lead to a low Z when the surface $\rm{Z/X}$ is fixed, thus the BCZ becomes shallow \citep{brun99,zha12}. However, the TKF could compensate it.

The required $\rm{L_{K,bc}}$ seems to be too high that is comparable with the total luminosity. According to \citet{tian09} and \citet{hotta14}, that is plausible. Xiong's turbulent convection model gives $\rm{L_{K,bc}} \sim -1\% L_{\odot}$ \citep{xiong01}. However, \citet{tian09} have shown that the gradient type model of the TKF in Xiong's model is too imprecise to be acceptable.

A serious problem that the stellar models could be significantly affected is arising if the TKF is in true far away from ignorable. The possibly effects may be in H/He main-sequence stars with convective core and RGB/AGB stars with convective dredge up, since the variation of the boundary of the convective core/envelope changes the profile of chemical abundance in stellar interior.

\acknowledgments

Many thanks to two referees for constructive comments which improve the original manuscript.
This work is co-sponsored by the National Natural Science
Foundation of China (NSFC) through grant No. 11303087, the West Light Foundation of the Chinese Academy
of Sciences, the Science Foundation of Yunnan Observatory No. Y1ZX011007 \& Y3CZ051005 and the Chinese Academy
of Sciences under grant No. KJCX2-YW-T24.

\end{document}